\documentclass[10pt,emptycopyrightspace]{ewsn-proc}

\usepackage{subfigure}
\usepackage{booktabs}
\usepackage{hyperref}
\usepackage{graphicx}
\usepackage{balance}
\usepackage{comment}

\usepackage{pifont}

\usepackage{xspace}

\newcommand{\etal}{\textit{et al.}}

\newcommand{\ie}{\textit{i.e.,}~}

\newcommand{\one}{({\em i})\xspace}
\newcommand{\two}{({\em ii})\xspace}
\newcommand{\three}{({\em iii})\xspace}

\let\orgautoref\autoref
\renewcommand{\autoref}
{\def\sectionautorefname{Section}%
\def\subsectionautorefname{Section}%
\def\subsubsectionautorefname{Section}%
\orgautoref}

\makeatletter
\renewcommand{\paragraph}[1]{\vspace*{0.03in}\noindent{\bf #1.}\hspace{0.25ex \@plus1ex \@minus.2ex}}
\newcommand{\paragraphS}[1]{\vspace*{0.03in}\noindent{\bf #1}\hspace{0.25ex \@plus1ex \@minus.2ex}}
\makeatother

\usepackage{diagbox}
\usepackage{fontawesome5}
\usepackage{tikz}
\usepackage{array}
\usetikzlibrary{automata,arrows,arrows.meta,positioning,calc,patterns,decorations, decorations.markings,external,shapes.geometric}
\usepackage{cases}
\usepackage{multirow}
\usepackage{pgfplots}
\usepackage{pgfplotstable}
\usepgfplotslibrary{groupplots,colormaps,colorbrewer,statistics}
\usepackage{makecell}
\usepackage{arydshln}
\usepackage[absolute,showboxes]{textpos}

\numberofauthors{3}
\author{
\alignauthor Jos\'e \'Alamos \\
    \affaddr{HAW Hamburg} \\
    \email{\normalsize jose.alamos@haw-hamburg.de}
\alignauthor Thomas C. Schmidt \\
    \affaddr{HAW Hamburg} \\
    \email{\normalsize t.schmidt@haw-hamburg.de}
\alignauthor Matthias W\"ahlisch \\
    \affaddr{TU Dresden \& Barkhausen Institut}\\
    \email{\normalsize m.waehlisch@tu-dresden.de}
}

\title{Deterministic IPv6 Networking with DSME-LoRa: Deterministic IPv6 Networking in the\\ Low Power Wireless IoT with DSME-LoRa}
\title{LoRa between Peers/From Peer to Peer: Full Stack IPv6 Networking with\\ DSME-LoRa on Low Power IoT Nodes}
\title{Peer-to-Peer LoRa: Full Stack IPv6 Networking with\\ DSME-LoRa on Low Power IoT Nodes}
\title{6LoRa: Full Stack IPv6 Networking with\\ DSME-LoRa on Low Power IoT Nodes}

\begin{document}

\maketitle

\setlength{\TPHorizModule}{\textwidth}
\setlength{\TPVertModule}{\paperheight}
\TPMargin{5pt}
\begin{textblock}{1}(.1,0.01)
\noindent
\footnotesize
If you refer to this paper, please cite the peer-reviewed publication::
J. Alamos, T. C. Schmidt, M. W{\"a}hlisch.\\
6LoRa: Full Stack IPv6 Networking with DSME-LoRa on Low Power IoT Nodes.\\
\emph{Proc. of the 2023 INTERNATIONAL CONFERENCE ON EMBEDDED WIRELESS SYSTEMS
AND NETWORKS., 2023.}
\end{textblock}

\begin{abstract}
Long range wireless transmission techniques such as LoRa are
preferential candidates for a substantial class of IoT applications, as they avoid the complexity of multi-hop wireless
forwarding. The existing network solutions for LoRa, however, are not suitable for peer-to-peer 
communication, which is a key requirement for many IoT applications.
In this work, we propose a networking system -- 6LoRa, that enables IPv6
communication over LoRa.
We present a full stack system implementation on RIOT OS and evaluate the
system on a real testbed using realistic application scenarios with CoAP.
Our findings confirm that our approach outperforms existing solutions in
terms of transmission delay and packet reception ratio at comparable energy
consumption.
\end{abstract}




\section{Introduction}\label{sec:intro}

Proliferation of proprietary IoT devices has led a fragmentation of IoT communication
protocols, which cha\-llen\-ges interoperation and slows down innovation. To
address this problem, the Internet Engineering Task Force (IETF) has proposed
adaptation layers to run IPv6 on top of a variety of IoT technologies such as
IEEE 802.15.4 (6LoWPAN~\cite{RFC-4944}) and LPWAN (SCHC~\cite{RFC-9011}).

Beyond interoperation, IPv6 has two additional advantages. \one It utilizes
existing infrastructure, which is well-proven with many successful deployments;
\two it enables the utilization of state of the art IoT application protocols
such as the Constrained Application Protocol (CoAP)~\cite{RFC-7252}.

The LoRaWAN ~\cite{lorawan-spec-11} LPWAN specification defines a cloud-based Media
Access Control layer protocol on top of the wireless modulation LoRa, which is
known to achieve
long range transmissions range (km) at very low power (mJ). LoRaWAN has
attracted special attention in the industry and academia for its open specification and low hardware cost.

\begin{figure}
    \centering
    \input{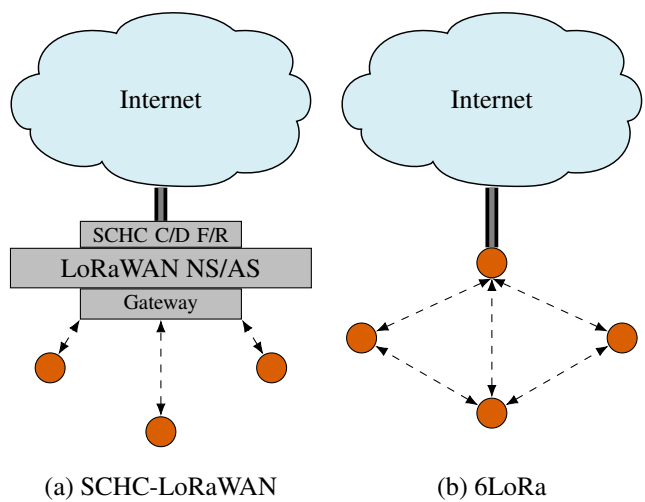}
    \caption{Network architecture of SCHC-LoRaWAN (left) and our solution (right).}
    \label{fig:lorawan_arch}
\end{figure}

While SCHC successfully enables transmission of IPv6 packets over LoRaWAN, its
centralized architecture (\autoref{fig:lorawan_arch}, left)
challenges applications that rely on downlink communication such as automation
control sce\-na\-rios. The architecture does not allow for direct peer-to-peer
communication between IoT devices, which is a common pattern in ad hoc networks. In consequence, the network must forward
messages from a transmitting device to the cloud and then back to the target device.
This approach exhibits three problems. \one Downlink communication consumes base
stations transmission budget and prevents concurrent reception of uplink
frames; \two it hinders the deployment of edge devices and leads to increased
costs in cloud infrastructure; \three in many remote areas, cellular networks
are the only feasible uplink option for the base stations, which renders
communication between IoT devices useless on poor Internet connectivity.


Recently, we proposed DSME-LoRa~\cite{aksw-dslrc-22}, an adaptation
layer to operate IEEE 802.15.4e Deterministic and Synchronous Multichannel
Extension (DSME)~\cite{IEEE-802.15.4-16} on top of the LoRa modulation. This
obsoletes the LoRaWAN network emulation. In this paper, we go one step further
and develop 6LoRa, a system to enable IPv6 communication over DSME-LoRa networks.
(\autoref{fig:lorawan_arch}, right).

The contributions of this paper are:

\begin{enumerate}
\item A system architecture for transmitting IPv6 frames over DSME-LoRa (6LoRa).
\item An implementation of 6LoRa in the RIOT operating system.
\item A comparative evaluation on real hardware of the system against SCHC-LoRaWAN.
\end{enumerate}

The remainder of this paper is structured as follows. We recap the  background 
with reviewing related work on LoRaWAN, IEEE 802.15.4 DSME and CoAP/IPv6 in
\autoref{sec:background}. \autoref{sec:system_design} introduces the system
design. \autoref{sec:implementation} describes the implementation in RIOT. We
show the results of our evaluation on real hardware in
\autoref{sec:evaluation} and conclude with an outlook in
\autoref{sec:conclusions}.

\section{Background and related work}\label{sec:background}

\paragraph{LoRaWAN}
LoRaWAN contributes a networking substrate to the long range transmission layer
of LoRa~\cite{lorawan-spec-11}.
Devices and cloud applications exchange data through a backhaul infrastructure
consisting of gateways (base stations), a network server (network controller),
and an application server (cloud services). Although communication is bidirectional, the architecture favors uplink communication.

LoRaWAN offers three communication classes, namely A, B and C, that trade-off
downlink latency and energy consumption. Class A seizes uplink packets for
polling downlink messages. Following packet transmission, devices open two reception windows
(RX1 and RX2) to receive a downlink packet. If the network server has multiple
packets to deliver, it sets the frame pending bit in the next downlink, which
prompts the receiving device to schedule an uplink as soon as possible. Class B
devices open reception windows at regular intervals. Class C devices keep the
reception window open all the time.

Mikhaylov~\etal~\cite{mpp-edtpl-18}
shows that downlink traffic compromises performance of uplinks, which hinders
deterministic communication. Although LoRaWAN class B decreases downlink latencies
compared to class A~\cite{ekbb-elcbe-20}, it suffers from scalability issues~\cite{fbf-eslgc-18,sak-lcbmc-20}. Vincenzo~\etal~\cite{vht-idsl-19} propose countermeasures for the limitations of LoRaWAN downlink traffic by adding multiple gateways
and providing a gateway selection mechanism. This increases deployment costs.
Zorbas~\etal~\cite{zakp-ttlii-20} present TS-LoRa,
which adds a time-slotted mechanism to LoRaWAN and achieves more than 99\% packet delivery ratio. Still, it does not address the downlink limitations of LoRaWAN.

\paragraph{IEEE 802.15.4 DSME}
The IEEE 802.15.4e Deterministic Synchronous Multichannel Extension (DSME)~\cite{IEEE-802.15.4-16}
defines a beacon-synchronized link layer. The MAC initiates a superframe structure,
which reserves time for contention-based (CSMA/CA)
and contention-free (GTS) transmission. The superframe structure repeats indefinitely.

CSMA/CA transmissions are carried out in a common channel, whereas GTS transmissions
are carried out in de\-di\-ca\-ted time-frequency slots. Devices
must negotiate a slot with the receiver prior GTS transmissions. The DSME MAC
allows for grouping superframes into a multisuperframe, which extends the
number of available GTS at the cost of increased transmission latency.

DSME devices may operate in two modes, namely coordinator and child.
Coordinators are responsible for network formation, maintenance and data
routing between children and devices beyond the radio range of each other.
The DSME MAC supports star, mesh and cluster-tree topologies depending on
the arrangement of coordinators and children. Among coordinators, one assumes a
significant role known as the PAN Coordinator, which dictates the superframe
structure for the entire network.

Battaglia~\etal~\cite{bcllp-neesr-20} present two novel extensions for the DSME
MAC, which address scalability issues in GTS handling in networks with a high
number of nodes and periodic flows. Vallati~\etal~\cite{vbpa-infid-17} discover
inefficiencies during DSME network formation and propose effective mitigations.
Alamos~\etal~\cite{aksw-dslrc-22} present and evaluate DSME-LoRa, which defines
an adaptation layer to interface the DSME MAC with a standard LoRa transceiver.
We employ the adaptation layer as a basis for our work.

\paragraph{IoT-centric IPv6 communication} SCHC (RFC 8724~\cite{RFC-8724}) describes a generic header compression and fragmentation of
IPv6 packets for transmission over a generic LPWAN.
It employs a set of out of band static rules (context) to instruct devices on
the procedure to compress and decompress IPv6 packets, selecting the most suitable rule.
RFC9011~\cite{RFC-9011} describes the LoRaWAN adaptation for SCHC.
The 6LoWPAN protocol~\cite{RFC-6282} enables the transmission of IPv6 packets
over IEEE 802.15.4 networks by using a set of headers (\ie IP Header Compression,
Next Header Compression for UDP) to facilitate compression and decompression,
fragmentation and reassembly of transmitted packets. The protocol supports both
stateful and stateless compression schemes.
In contrast to 6LoWPAN, the compression scheme of SCHC performs better~\cite{gmtbz-iolcl-20} and supports
compression of the CoAP header~\cite{RFC-8824}. This has motivated research efforts towards
the transmission of SCHC-compressed packets over IEEE 802.15.4 networks~\cite{draft-gomez-6lo-schc-15dot4-03}.

The Constrained Application Protocol~\cite{RFC-7252} (CoAP) is a lightweight
Web protocol based on UDP. The protocol follows a Request/Response model and
incorporates REST-like methods, including GET, POST and PUT. 

\section{System design}\label{sec:system_design}

\begin{figure}
    \centering
\begin{tikzpicture}[node distance=1.2cm and 0.3cm]
\input{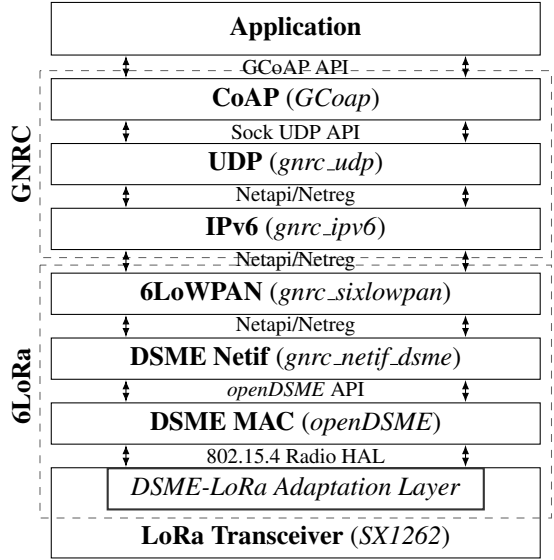}
\tikzstyle{app} = [rectangle, minimum width=6.5cm, minimum height=0.7cm, text centered, draw=black]
\tikzstyle{small} = [rectangle, minimum width=4cm, minimum height=0.4cm, text centered, draw=black]
\tikzstyle{big} = [rectangle, minimum width=6.5cm, minimum height=0.4cm, text centered, draw=black]
\tikzstyle{line} = [thick,-,>=stealth]
\tikzstyle{arrow} = [thick,->,>=stealth]
\tikzstyle{pr} = [xshift=1cm]
\tikzstyle{pl} = [xshift=-1cm]

\node (app) [app] {\textbf{Application}};
\node (gcoap) [big,below=0.3cm of app.south east,anchor=north east] {\textbf{CoAP} (\textit{GCoap})};
\node (gnrc_udp) [big,below=0.3cm of gcoap] {\textbf{UDP} (\textit{gnrc\_udp})};
\node (gnrc_ipv6) [big,below=0.3cm of gnrc_udp] {\textbf{IPv6} (\textit{gnrc\_ipv6})};
\node (gnrc_sixlowpan) [big,below=0.3cm of gnrc_ipv6] {\textbf{6LoWPAN} (\textit{gnrc\_sixlowpan})};
\node (netif_opendsme) [big,below=0.3cm of gnrc_sixlowpan.south east, anchor=north east] {\textbf{DSME Netif} (\textit{gnrc\_netif\_dsme})};
\node (mac) [big,below=0.3cm of netif_opendsme] {\textbf{DSME MAC} (\textit{openDSME})};
\node (loraphy) [app,minimum height=1.2cm, below=0.3cm of mac] {};
\node (dsmelora) [big,minimum height=0.4cm, minimum width= 5cm, below=0cm of loraphy.north, anchor=north,thick,draw=black!80!white] {\textit{DSME-LoRa Adaptation Layer}};
\node[anchor=south] at (loraphy.south) {\textbf{LoRa Transceiver} (\textit{SX1262})};

\draw[dashed,black!60!white] ([xshift=-3.4cm,yshift=0.1cm]gnrc_sixlowpan.north) rectangle ([xshift=3.4cm,yshift=-0.1cm]dsmelora.south);
\node[rotate=90,anchor=south] at ([xshift=-3.4cm,yshift=0.5cm] mac) {\textbf{6LoRa}};

\draw[dashed,black!60!white] ([xshift=-3.4cm,yshift=0.1cm]gcoap.north) rectangle ([xshift=3.4cm,yshift=-0.1cm]gnrc_ipv6.south);
\node[rotate=90,anchor=south] at ([xshift=-3.4cm] gnrc_udp) {\textbf{GNRC}};

\draw [<->,>={[scale=0.75]Latex}] ([pr]gcoap.north west) -- ([pr]gcoap.north west |- app.south);
\draw [<->,>={[scale=0.75]Latex}] ([pr]gnrc_udp.north west) -- ([pr]gnrc_udp.north west |- gcoap.south);
\draw [<->,>={[scale=0.75]Latex}] ([pr]gnrc_ipv6.north west) -- ([pr]gnrc_ipv6.north west |- gnrc_udp.south);
\draw [<->,>={[scale=0.75]Latex}] ([pr]gnrc_sixlowpan.north west) -- ([pr]gnrc_sixlowpan.north west |- gnrc_ipv6.south);
\draw [<->,>={[scale=0.75]Latex}] ([pr]gnrc_sixlowpan.south west) -- ([pr]gnrc_sixlowpan.south west |- netif_opendsme.north west);
\draw [<->,>={[scale=0.75]Latex}] ([pr]mac.north west) -- ([pr]mac.north west |- netif_opendsme.south);
\draw [<->,>={[scale=0.75]Latex}] ([pr]loraphy.north west) -- ([pr]loraphy.north west |- mac.south);


\draw [<->,>={[scale=0.75]Latex}] ([pl]gcoap.north east) -- ([pl]gcoap.north east |- app.south);
\draw [<->,>={[scale=0.75]Latex}] ([pl]gnrc_udp.north east) -- ([pl]gnrc_udp.north east |- gcoap.south);
\draw [<->,>={[scale=0.75]Latex}] ([pl]gnrc_ipv6.north east) -- ([pl]gnrc_ipv6.north east |- gnrc_udp.south);
\draw [<->,>={[scale=0.75]Latex}] ([pl]gnrc_sixlowpan.north east) -- ([pl]gnrc_sixlowpan.north east |- gnrc_ipv6.south);
\draw [<->,>={[scale=0.75]Latex}] ([pl]gnrc_sixlowpan.south east) -- ([pl]gnrc_sixlowpan.south east |- netif_opendsme.north east);
\draw [<->,>={[scale=0.75]Latex}] ([pl]mac.north east) -- ([pl]mac.north east |- netif_opendsme.south);
\draw [<->,>={[scale=0.75]Latex}] ([pl]loraphy.north east) -- ([pl]loraphy.north east |- mac.south);

\path [] (gcoap.north) -- (gcoap.north |- app.south) node[pos=0.5] {\footnotesize GCoAP API};
\path [] (gnrc_udp.north) -- (gnrc_udp.north |- gcoap.south) node[pos=0.5] {\footnotesize Sock UDP API};
\path [] (gnrc_ipv6.north) -- (gnrc_ipv6.north |- gnrc_udp.south) node[pos=0.5] {\footnotesize Netapi/Netreg};
\path [] (gnrc_sixlowpan.north) -- (gnrc_sixlowpan.north |- gnrc_ipv6.south) node[pos=0.5] {\footnotesize Netapi/Netreg};
\path [] (netif_opendsme.north) -- (netif_opendsme.north |- gnrc_sixlowpan.south) node[pos=0.5] {\footnotesize Netapi/Netreg};
\path [] (mac.north) -- (mac.north |- netif_opendsme.south) node[pos=0.5] {\footnotesize \textit{openDSME} API};
\path [] (loraphy.north) -- (loraphy.north |- mac.south) node[pos=0.5] {\footnotesize 802.15.4 Radio HAL};

%
%
%

\end{tikzpicture}
    \caption{System architecture of a 6LoRa network stack. The figure includes software components and interfaces of the implementation in the RIOT operating system.}
    \label{fig:system_design}
\end{figure}

\autoref{fig:system_design} summarizes the system architecture for a 6LoRa network stack.
The design caters for IPv6 packet transmission over LoRa.

We utilize the DSME MAC to enable reliable communication of IPv6 packets in 
LoRa frames. The MAC interacts with the LoRa transceiver via the DSME-LoRa adaptation layer~\cite{aksw-dslrc-22}.
The DSME network interface grants full control to the upper layer of the transmission pattern (CSMA/CA or
slotted), the superframe duration and slot allocation.
To enable compression and fragmentation of IPv6 packets, we use the 6LoWPAN~\cite{RFC-6282} adaptation layer.
This is required to reduce link stress and decrease time on air of LoRa frames. 
On top of 6LoRa, we
consider a standard IoT network stack consisting of CoAP via UDP and IPv6.
Similar to traditional
6LoWPAN networks, devices in the network may act as routers to forward traffic between
devices or to the Internet (not evaluated in this work).

\section{Implementation}\label{sec:implementation}

We extend the implementation of DSME-LoRa in the RIOT operating system~\cite{bghkl-rosos-18}
to enable IPv6 support via the network stack (GNRC), as shown
in \autoref{fig:system_design}. The implementation provides DSME support through
the \textit{openDSME} package and implements the \textit{802.15.4 Radio HAL}
interface with the DSME-LoRa adaptation layer, in order to interface the DSME
MAC with the LoRa transceiver (\textit{SX1262}).
The GNRC network stack runs each protocol layer  on a dedicated thread and uses
inter process communication via a common API (GNRC Netapi/Netreg) to communicate between layers.
This approach facilitates protocol implementation.

We add IPv6 support to the existing DSME-LoRa network interface
(\textit{gnrc\_netif\_dsme}).
For that, we take care of three aspects. \one the initialization of the
necessary IPv6 Information Base, such as hop limit and network address; \two
the configuration of the interface for use with 6LoWPAN and \three perform
broadcast transmission for packets which where already tagged by the
network stack as multicast.


The 6LoRa network stack utilizes the UDP (\texttt{gnrc\_udp}) and IPv6
(\texttt{gnrc\_ipv6}) components of the network stack for the transmission of
CoAP frames and 6LoWPAN (\texttt{gnrc\_sixlowpan}) to transmit compressed IPv6 packets
over the DSME MAC.

GNRC implements the Sock UDP API (\texttt{sock\_udp}), which exposes network
stack agnostic UDP sockets. We utilize the RIOT CoAP implementation (GCoAP),
which uses the Sock UDP API to send and receive CoAP packets.

\section{Evaluation}\label{sec:evaluation}

We assess the performance of CoAP communication over SCHC-LoRaWAN and 6LoRa devices.
We also compare memory requirements and power consumption of both stacks. In
this study, fragmentation is not evaluated.


\begin{figure}
    \centering
    \subfigure[SCHC-LoRaWAN] {
        \label{fig:ev_schc}
        \definecolor{color0}{rgb}{0.905882352941176,0.16078431372549,0.541176470588235}
\definecolor{color1}{rgb}{0.850980392156863,0.372549019607843,0.00784313725490196}
\definecolor{color2}{rgb}{0.458823529411765,0.43921568627451,0.701960784313725}
\definecolor{color3}{rgb}{0.105882352941176,0.619607843137255,0.466666666666667}
\definecolor{color4}{rgb}{0.4,0.650980392156863,0.117647058823529}
\begin{tikzpicture}[>={Latex}]
  \tikzset{
  bignode/.style = {draw, fill=white, rectangle, inner sep=3pt, minimum width=3cm, minimum height=0.9cm},
    inner/.style = {draw, fill=white, rectangle, inner sep=2pt, minimum width=1.8cm, font=\small},
    innerbottom/.style = {inner, anchor=south},
    innertop/.style = {inner, anchor=north},
    sensor/.style = {draw, circle, inner sep=2pt },
  }

  \node[bignode] (linux) {};
    \node[anchor=north] at (linux.north) {Linux machine};
    \node[innerbottom] at (linux.south) {\textit{pylibschc}};
    \node[bignode, below= 0.7cm of linux] (gw) {};
    \node[anchor=south] at (gw.south) {Gateway};
    \node[innertop]  at (gw.north) {\textit{Chirpstack}};

  \draw[black, double=gray,double distance=2pt,line width=0.05cm]
    ([xshift=0.7cm]gw.north) -- ([xshift=0.7cm]linux.south) node[midway, anchor=west,font=\small] {MQTT};

        \node[sensor, below=0.7cm of gw] (c1) {\Large \faLightbulb[regular]};
        \node[sensor] (c2) at ([shift={(-30:1.5cm)}]c1) {\Large \faThermometerHalf};
        \node[sensor] (c3) at ([shift={(-150:1.5cm)}]c1) {\Large \faThermometerHalf };

        \node[anchor=north] at (c1.south) {C1};
        \node[anchor=north] at (c2.south) {C2};
        \node[anchor=north] at (c3.south) {C3};
    
      \draw[->]
        (c2) -- ([xshift=-0.2cm]gw.south east)
        ;
      \draw[->]
        ([xshift=-0.2cm]gw.north) -- 
        ([xshift=-0.2cm]linux.south);
        ;
    
      \draw[<-]
        ([xshift=0.2cm]gw.north) -- ([xshift=0.2cm]linux.south);
    
      \draw[<-]
        (c1) -- (gw.south)
        ;
      \draw[->]
        (c3) -- ([xshift=0.2cm]gw.south west)
        ;

\end{tikzpicture}
    }
    \subfigure[6LoRa] {
        \definecolor{color0}{rgb}{0.905882352941176,0.16078431372549,0.541176470588235}
\definecolor{color1}{rgb}{0.850980392156863,0.372549019607843,0.00784313725490196}
\definecolor{color2}{rgb}{0.458823529411765,0.43921568627451,0.701960784313725}
\definecolor{color3}{rgb}{0.105882352941176,0.619607843137255,0.466666666666667}
\definecolor{color4}{rgb}{0.4,0.650980392156863,0.117647058823529}
\begin{tikzpicture}[>={Latex}]
  \tikzset{
  bignode/.style = {draw, fill=white, rectangle, inner sep=3pt, minimum width=3cm, minimum height=0.9cm},
    inner/.style = {draw, fill=white, rectangle, inner sep=2pt, minimum width=1.8cm, font=\small},
    innerbottom/.style = {inner, anchor=south},
    innertop/.style = {inner, anchor=north},
    sensor/.style = {draw, circle, inner sep=2pt },
  }


        \node[sensor] (c1) {\Large \faLightbulb[regular]};
        \node[sensor] (c2) at ([xshift=1.2cm,yshift=-2.5cm]c1) {\Large \faThermometerHalf};
        \node[sensor] (c3) at ([xshift=-1.2cm, yshift=-2.5cm]c1) {\Large \faThermometerHalf };

        \node[anchor=south] at (c1.north) {C1};
        \node[anchor=north] at (c2.south) {C2};
        \node[anchor=north] at (c3.south) {C3};
    
      \draw[->]
        (c3) -- (c1)
        ;
      \draw[->]
        (c2) -- (c1)
        ;

\end{tikzpicture}
        \label{fig:ev_dsme}
    }
    \label{fig:evaluation}
    \caption{Network topology and traffic flow of the SCHC-LoRaWAN network (left) and
    6LoRa network (right) used in our evaluation. C1 corresponds to the cluster with three
    actuators (\faLightbulb[regular]), while C2 and C3 represent clusters with eight and four sensors (\faThermometerHalf), respectively.}
\end{figure}
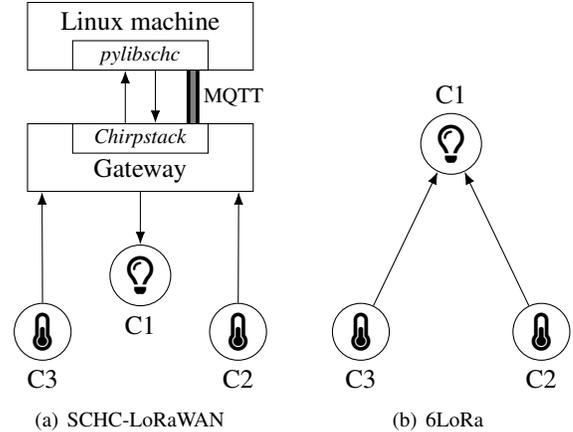

\subsection{Experimental setup}
Experiments are conducted using commercial off-the-shelf platforms (Nucleo-WL55JC1).
The platform is equipped with a dual core ARM
Cortex-M4F and ARM Cortex-M0+ CPU, with a clock speed of 48\,MHz and a me\-mo\-ry
capacity of 256\,kB of ROM/64\,kB of RAM. Due to the absence of dual-core
support in RIOT, we solely utilize the Cortex-M4F core. The platform is further
equipped with a LoRa transceiver (SX1262).

Fifteen devices were deployed in the university facilities, grouped in three clusters.
The first cluster (C1) consists of three LoRa devices that function as actuator
devices. The second
cluster (C2) includes eight LoRa devices and is located on the floor above
C1, approximately 60\,m away of C1. The third cluster (C3) is composed of four
LoRa devices and is deployed without line-of-sight on the same floor of C1, approximately 12\,m
away. Devices in C2 an C3 operate as sensor devices.

Each sensor device from C2 and C3 sends 12 bytes in non-confirmed CoAP POST
packets to an actuator from C1. Out of the twelve sensors,
six of them send data exclusively to the first actuator. Among the remaining sensors,
three of them send data to the second actuator, while the other three send data to the
third actuator.
To minimize wireless traffic, we configure CoAP to elide responses.
The transmission intervals are governed by a
uniform distribution characterized by means of 10\,s and 20\,s, and a range of
plus or minus 2 seconds. The LoRaWAN infrastructure 
forwards traffic between end devices and the SCHC endpoint
(\autoref{fig:ev_schc}), while 6LoRa devices employ direct
peer-to-peer communication (\autoref{fig:ev_dsme}).

We deploy a RAK2245 LoRaWAN gateway next to the
C1 cluster, which runs the \textit{Chirpstack} LoRaWAN network server (v3.15.1)
and application server (v3.17.2). The network server is configured to use all
eight LoRaWAN channels that the gateway supports. In order to minimize the
time on air of downlink packets, the datarate of the second reception
window (RX2) channel is set to 5 (SF7BW125). The evaluation does not consider class
B due to limited support in RIOT.

We employ the \textit{libschc} library~\cite{mkhpj-usfop-19} for the SCHC compression
and decompression in the constrained devices. Additionally, we utilize a simple Python
implementation that uses \textit{pylibschc}, a \textit{libschc} wrapper,
to communicate with Chirpstack via MQTT and expose a SCHC endpoint on a Linux machine. 

We configure a single SCHC rule for the compression of IPv6 and UDP headers, that
achieves the maximum level of compression attainable by SCHC for our experiment.
Due to its incomplete CoAP support, \textit{libschc} does not compress the CoAP header.

SCHC-LoRaWAN actuator devices operating under class A protocol employ 
polling packets with empty payload at a regular 10-second interval, in order to retrieve
downlink packets from the network server.

In 6LoRa scenarios we evaluate only GTS transmissions. The choice is
made to mitigate the detrimental effects of concurrent channel access in CSMA/CA, which
degrades packet reception~\cite{aksw-dslrc-22}. We set the superframe
duration to the minimum feasible time (3.84\,s) that allows transmissions of
the CoAP POST packet within a GTS. Devices allocate only one GTS, which repeats at
the superframe duration (3.84\,s).

SCHC-LoRaWAN and 6LoRa devices do not request acknowledgements and transmit using
the same PHY configuration (SF7BW125).
For 6LoWPAN we use stateful, context-based compression for the IPv6 source and
destination addresses.


\subsection{Memory requirements}

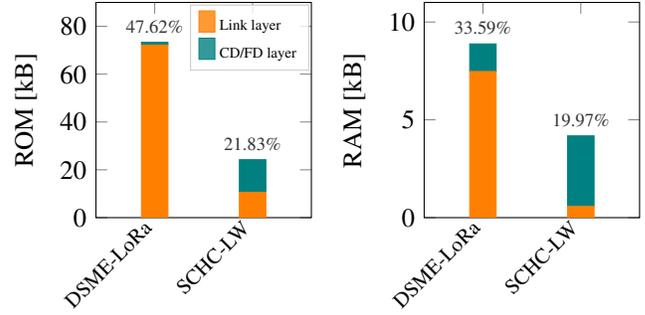
\begin{figure}
    \centering
    \begin{center}
\begin{tikzpicture}
\tikzset{
  txt/.style = {font=\scriptsize,color=black!80!white},
}
\begin{groupplot}[
ymin=0,
ymax=90,
group style={
  group size=2 by 1,
  horizontal sep=1.5cm,
  vertical sep=0.35cm,
},
width=0.25\textwidth,
height=0.25\textwidth,
symbolic x coords={
{DSME-LoRa},
{SCHC-LW},
},
legend cell align={left},
legend style={
legend columns=1,
fill opacity=0.8,
draw opacity=1,
text opacity=1,
font=\scriptsize,
at={(0.99,0.99)},
nodes={scale=0.8, transform shape},
anchor=north east,
draw=white!80!black,
},
xtick=data,
xticklabel style={rotate=45,anchor=east,font=\footnotesize},
ylabel={Memory (kB)},
ylabel style={yshift=-0.3cm},
enlarge x limits=0.6,
ybar stacked,
]
\nextgroupplot[ylabel={ROM [kB]}]
\addplot+ [fill=orange,draw=orange
] coordinates {
({DSME-LoRa}, 72.19)
({SCHC-LW}, 10.61)
};
\addlegendentry{Link layer}
\addplot+ [fill=teal,draw=teal
] coordinates {
({DSME-LoRa}, 1.12)
({SCHC-LW}, 13.62)
};
\node[above,txt] at (axis cs:SCHC-LW,24.23) {21.83\%};
\node[above,txt] at (axis cs:DSME-LoRa,73.31) {47.62\%};
\addlegendentry{CD/FD layer}
\nextgroupplot[ymax=11, ylabel={RAM [kB]}]
\addplot+ [fill=orange,draw=orange
] coordinates {
({DSME-LoRa}, 7.47)
({SCHC-LW}, 0.58)
};
\addplot+ [fill=teal,draw=teal
] coordinates {
({DSME-LoRa}, 1.41)
({SCHC-LW}, 3.60)
};
\node[above,txt] at (axis cs:SCHC-LW,4.18) {19.97\%};
\node[above,txt] at (axis cs:DSME-LoRa,8.88) {33.59\%};
\end{groupplot}
\end{tikzpicture}
\end{center}
    \caption{ROM (left) and RAM (right) requirements for SCHC-LoRaWAN and
    6LoRa in RIOT, separated in compression/decompresion
    fragmentation/reassembly (CD/FR) and link layer, measured on a
    Nucleo-WL55JC1 platform (ARM Cortex-M4F). Relative values relate to the
    total firmware size.}
    \label{fig:firmware_size}
\end{figure}

\autoref{fig:firmware_size} shows the memory requirements, separated into ROM
(\textit{text} + \textit{data} segment) and RAM (\textit{data} + \textit{bss}
segment), for SCHC-LoRaWAN and 6LoRa in our target platform
(Nucleo-WL55JC1).

We evaluate only the memory requirements for the compression/decompression
fragmentation/reassembly layer (SCHC, 6LoWPAN) and link layer (LoRaWAN, DSME-LoRa).
Hence, we do not consider the memory requirements for components of the operating system (\ie GNRC
network stack, RIOT core and LoRa driver), as these components do
not vary significantly between the two firmwares.

The SCHC layer requires 13.62\,kB of ROM and 3.60\,kB of RAM. This layer includes the \textit{libschc} library, which stores internal
buffers in RAM, and the RIOT contrib code, which allocates $\approx$ 1\,kB in
RAM for the \textit{libschc} thread stack.

The LoRaWAN stack (GNRC LoRaWAN) requires 10.61\,kB of ROM.
The stack operates frugal and requires only 0.58\,kB of RAM to store
the MAC descriptor and the MAC Information Base. 

To summarize, the SCHC-LoRaWAN implementation utilizes a total of 24.23\,kB (21.83\%) of ROM and 4.18\,kB (19,97\%) of RAM.

The 6LoWPAN implementation (GNRC 6LoWPAN) requires 1.12\,kB of ROM for the protocol logic and 1.41\,kB of RAM
for the thread stack. 

The DSME-LoRa implementation includes \textit{openDSME}, the contrib code for the platform
abstraction and the DSME-LoRa Adaptation Layer. The implementation requires 72.19\,kB of ROM and 7.47\,kB of RAM, primarily attributed to the DSME MAC
implementation. This MAC is more intricate compared to the LoRaWAN MAC and necessitates additional resources for storing
internal data structures. To summarize, 6LoRa requires 73.31\,kB (47.62\%) of ROM and 8.88\,kB (33.59\%) of RAM.

\begin{figure}
    \centering
    \tikzexternalenable%
\tikzsetnextfilename{tx_delay}
\begin{tikzpicture}
\tikzset{
  prr/.style = {draw=black!80!white,dashed},
  uplink/.style = {draw=red!80!white},
}
\pgfplotstableread[]{data/lw_a_post_10.dat}\LWAten
\pgfplotstableread[]{data/lw_a_post_20.dat}\LWAtwenty
\pgfplotstableread[]{data/lw_post_10.dat}\LWten
\pgfplotstableread[]{data/lw_post_20.dat}\LWtwenty
\pgfplotstableread[]{data/dsme_post_10.dat}\DSMEten
\pgfplotstableread[]{data/dsme_post_20.dat}\DSMEtwenty

\pgfplotstablecreatecol[
  create col/expr={0.767*\thisrowno{1}},
]{ScaledY}\LWAtwenty
\pgfplotstablecreatecol[
  create col/expr={0.852*\thisrowno{1}},
]{ScaledY}\LWtwenty
\pgfplotstablecreatecol[
  create col/expr={0.996*\thisrowno{1}},
]{ScaledY}\DSMEtwenty

\pgfplotstablecreatecol[
  create col/expr={0.732*\thisrowno{1}},
]{ScaledY}\LWAten
\pgfplotstablecreatecol[
  create col/expr={0.79*\thisrowno{1}},
]{ScaledY}\LWten
\pgfplotstablecreatecol[
  create col/expr={1*\thisrowno{1}},
]{ScaledY}\DSMEten

\begin{groupplot}[
group style={
  group size=3 by 2,
  horizontal sep=0.15cm,
  vertical sep=0.6cm,
},
height=0.20\textwidth,
width=0.20\textwidth,
ymax=1.1,
axis on top,
yticklabel pos=left,
xmin=0, ymin=0,
legend cell align={left},
legend style={
  fill opacity=0.8,
  draw opacity=1,
  text opacity=1,
  at={(0.97,0.03)},
  nodes={scale=0.8, transform shape},
  anchor=south east,
  draw=white!80!black,
},
tick align=outside,
tick pos=left,
x grid style={white!69.0196078431373!black},
xtick style={color=black},
y grid style={white!69.0196078431373!black},
ytick style={color=black},
cycle list name=exotic,
ytick={0,0.25,0.5,0.75,1},
]
\nextgroupplot[title=\textbf{Class A},xmax=39]
\addplot+ [mark repeat=80] table [x index=0, y index=2] {\LWAtwenty};
\addplot[forget plot, prr] coordinates {(0,0.767) (60,0.767)};
\addplot[forget plot, uplink] coordinates {(0,0.833) (60,0.833)};
\node[anchor=south west,font=\footnotesize] at (axis cs: 0,0.833) {$\Delta=6.6\%$};
\nextgroupplot[title={\textbf{Class C}}, xmax=13, yticklabels={}]
\addplot+ [mark repeat=80] table [x index=0, y index=2] {\LWtwenty};
\addplot[forget plot, prr] coordinates {(0,0.852) (60,0.852)};
\addplot[forget plot, uplink] coordinates {(0,0.855) (60,0.855)};
\node[anchor=south west,font=\footnotesize] at (axis cs: 0,0.855) {$\Delta=0.33\%$};
\nextgroupplot[title={\textbf{6LoRa}}, xmax=4, yticklabels={}]
\addplot+ [mark repeat=80] table [x index=0, y index=2] {\DSMEtwenty};
\addplot[forget plot, prr] coordinates {(0,0.996) (60,0.996)};
\nextgroupplot[xmax=39]
\addplot+ [mark repeat=80] table [x index=0, y index=2] {\LWAten};
\addplot[forget plot, prr] coordinates {(0,0.732) (60,0.732)};
\addplot[forget plot, uplink] coordinates {(0,0.74) (60,0.74)};
\node[anchor=south west,font=\footnotesize] at (axis cs: 0,0.74) {$\Delta=0.83\%$};
\nextgroupplot[xmax=59, yticklabels={}]
\addplot+ [mark repeat=80] table [x index=0, y index=2] {\LWten};
\addplot[forget plot, prr] coordinates {(0,0.79) (60,0.79)};
\addplot[forget plot, uplink] coordinates {(0,0.798) (60,0.798)};
\node[anchor=south west,font=\footnotesize] at (axis cs: 0,0.798) {$\Delta=0.5\%$};
\nextgroupplot[xmax=4, yticklabels={}]
\addplot+ [mark repeat=80] table [x index=0, y index=2] {\DSMEten};
\addplot[forget plot, prr] coordinates {(0,1) (60,1)};

\end{groupplot}
\node[anchor=base,rotate=-90,yshift=0.275cm] at (group c3r1.east) {\textbf{TX int=20\,s}};
\node[anchor=base,rotate=-90,yshift=0.275cm] at (group c3r2.east) {\textbf{TX int=10\,s}};
\node[anchor=base,rotate=90,yshift=1cm] at (group c1r1.south west) {CDF};
\node[anchor=base,rotate=0,yshift=-1cm] at (group c2r2.south) {Completion time [s]};

\end{tikzpicture}
\tikzexternaldisable%
    \caption{Distribution of completion time for transmitted CoAP packets over SCHC-LoRaWAN (class A and C) and 6LoRa, for varying transmission interval (20\,s and 10\,s).
    The intersection between curve and right axis relates to the packet
    reception ratio (dashed line). Data extraction ratio (\ie successful uplink packets) is shown for
    LoRaWAN (red line). Difference between red and dashed lines ($\Delta$) relates to downlink losses.}
    \label{fig:tx_delay}
\end{figure}
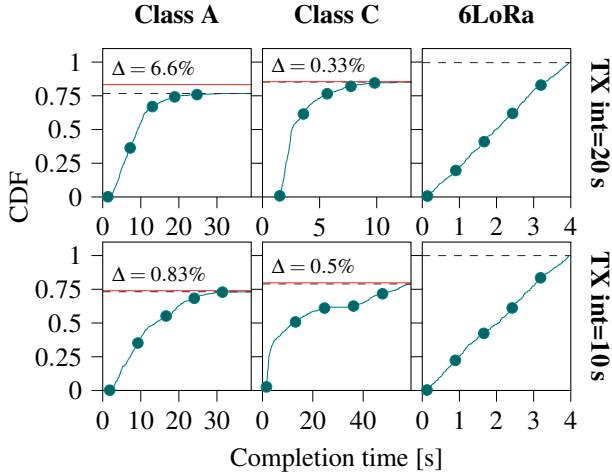

\subsection{Communication performance}\label{sec:performance}

In both scenarios, class C demonstrates a greater packet reception ratio compared to
class A. Specifically, class C achieves an 85\% ratio in the relaxed scenario (\autoref{fig:tx_delay}, top center) and an 80\%
ratio in the stressed scenario (\autoref{fig:tx_delay},bottom center), while class A achieves a 77\% ratio in the relaxed scenario (\autoref{fig:tx_delay},top left) 
and a 73\% ratio in the stressed scenario (\autoref{fig:tx_delay}, bottom left).
The better performance of class C occurs for two reasons: \one polling packets from class A actuators
increases on-air traffic and consequently degrades channel quality, ultimately
reducing data extraction ratio, and \two
a fraction of class A downlink packets are scheduled in the first reception window (RX1),
which utilizes the same frequency and datarate as uplink transmissions. As a result,
uplink and downlink frames collide~\cite{sgiv-udano-22}. This does not occur for class C devices,
as they receive most downlink packets in the same frequency of the second reception
window (RX2), which uses a frequency that does not interfere with uplink transmissions.

A shorter transmission interval (10\,s)
decreases data extraction ratio (\ie successfully received uplink packets) for three reasons: \one the increased on-air traffic from sensors
increases uplink collisions; \two
higher downlink traffic increases transmission duty cycle at the gateway,
which reduces its availability to receive uplink traffic; and \three the
network server frequently sets the frame pending bit for class A devices, leading to
increased polling packet traffic and degraded uplink reception.
Observe that class A devices (\autoref{fig:tx_delay}, left) experience higher downlink losses in the relaxed scenario
(6.6\%) than in the
stressed scenario (0.83\%). This outcome,
which may appear counterintuitive, is attributed to the preference of the network server
for the first reception window (RX1) for downlink traffic, which results in collisions
with uplink traffic. In the stressed scenario, the network server
schedules downlink traffic more often in the second reception window (RX2), whose traffic
does not collide with uplink traffic.

The completion time is given by the sum of uplink transmission time
($\approx$ 260\,ms, measured from the beginning of the transmission until the
reception on the SCHC endpoint), the SCHC endpoint processing time ($\approx$ 5\,ms),
the downlink schedule delay (variable) and the downlink transmission time
($\approx$ 1.3\,s, which includes the gateway delay and the transmission time on
air of the downlink packet).

The downlink schedule delay, which determines the completion time, depends on the downlink throughput and the downlink queue stress.
Class A throttles downlink rate to the polling rate of the receiving device, whereas class C
throttles the downlink rate to one downlink message every $\approx$ 2\,s.
As a result, even under low queue stress (\ie relaxed scenario,
\autoref{fig:tx_delay}, top), class C exhibits lower
completion time (less than 7\,s for 95\% of packets), than class A
(less than $\approx$ 17\,s for 95\% of packets).

In the stressed scenario, the downlink queue experiences high stress due to
increased sensor traffic. For class A actuators, this influx of traffic has an impact
on the delivery of polling packets.
These combined effects lead to extended waiting times for packets
in the downlink queue, ultimately resulting in a significant increase in completion time.
The knee point at approximately 35\,s for class C devices in the
stressed scenario (\autoref{fig:tx_delay}, bottom center) can be
attributed to a continuous build up of downlink packets for the
actuator with six sensor devices, which schedules in average at a shorter interval
($\approx$ 1.67) than the throttling interval (2\,s).

The completion time of 6LoRa transmissions is
primarily determined by the transmission delay of the DSME MAC layer, which is
influenced by the GTS schedule, and the time on air of the packet ($\approx$
118\,ms). When the application schedules a transmission
just before the occurrence of the next slot, the lowest completion time is
achieved, resulting in less than 140\,ms (\autoref{fig:tx_delay}, right).
Conversely, if the application schedules the transmission immediately after the
occurrence of the next slot, the packet is held in the MAC queue until the next
slot occurs (3.84\,s after the last slot), resulting in a worst-case completion
time of 100\% of packets transmitted in less than $\approx$ 3.9\,s. The
completion time remains constant in both the relaxed and stressed scenarios
because the inter-packet gap is always greater than the superframe duration.
Consequently, all packets are scheduled when the MAC queue is empty, which effectively
bounds the completion time to a value that aligns with the superframe duration.

In contrast to LoRaWAN traffic, DSME-LoRa traffic is not vulnerable to
collisions thanks to the time-division multiple access scheme (TDMA) of
GTS, which enables collision-free transmissions. The empirical results
indicate a significant packet reception rate, reaching 99.7\% for the relaxed
scenario and 100\% for the stressed scenario.

\subsection{Energy consumption}

We utilize a digital multimeter (DMM7510 7 1/2) to measure the current at a sampling rate of 100\,kHz
in order to estimate the power consumption of the devices.
Our analysis focused on the power consumption of sensors
as well as actuators that receive data from three sensors each.
To conserve energy, the 6LoRa devices in our analysis disable the transceiver during the contention
access period (reserved for CSMA/CA transmissions).

\begin{table}
\centering
    \caption{Power consumption of SCHC-LoRaWAN (class A and C) , and 6LoRa devices, for sensor devices with transmission intervals (\textbf{TXi}) of 20\,s and 10\,s, and for actuator devices.}
    \begin{tabular}{crrrr}
        \toprule
        \textbf{Device} & \textbf{TXi [s]} & \multicolumn{3}{c}{\textbf{Power [mW]}} \\
        &&\multicolumn{2}{c}{{SCHC-LoRaWAN}}&6LoRa \\
        \cmidrule(lr){3-4}
        && \textit{Class A} & \textit{Class C} & \\
        \midrule
        Sensor & 20 & 0.49 & 12.87 & 1.33 \\
        Sensor & 10 & 0.87 & 13.3 & 2.04 \\
        Actuator & - & 0.54 & 12.41 & 2.93 \\
        \bottomrule
    \end{tabular}
\label{tab:energy}
\end{table}

\autoref{tab:energy} illustrates the power consumption (over base consumption)
for SCHC-LoRaWAN and 6LoRa devices.

Unlike the DSME MAC, LoRaWAN class A and class C devices have a low
maintenance overhead since the MAC does not require maintaining a continuous connection with the network
server. Additionally, class A devices keep the transceiver off most of the
time, resulting in the lowest maintenance overhead and consequently, the lowest
power consumption across all scenarios. Although 6LoRa
devices yield low transceiver duty cycle, the DSME MAC consumes energy in order to
maintain the superframe structure (\ie synchronize to beacons and manage GTS).
This results in a higher maintenance overhead for 6LoRa devices, which
yields a higher power consumption than class A
devices in all scenarios but does not prevent operation on batteries. Class C
devices show the highest power consumption, as a result of the transceiver
listening continuously.

The power consumption of SCHC-LoRaWAN sensors increases by $\approx$ 0.4\,mW for the
stressed scenario (10\,s), whereas the power consumption of 6LoRa sensors
increases by $\approx$ 0.7\,mW. This is attributed to \one the shorter time on
air (108\,ms) of the LoRaWAN packet as a result of the superior compression scheme of SCHC
and \two inefficient handling in \textit{openDSME} of the GTS schedule, where the
transceiver is turned on too early. For this reason, the energy consumption
per transmitted packet is lower for SCHC-LoRaWAN devices than for 6LoRa devices.

Receiving class A devices transmit polling packets (10\,s transmission interval)
which consumes $\approx$ 0.54\,mW. Unlike
data packets, polling packets do not carry a payload, resulting in a lower
time on air and lower power consumption compared to class A sensor devices with 10\,s transmission
interval ($\approx$ 0.87\,mW). Class C devices consume $\approx$ 12.41\,mW as
a result of the continuous listening of the transceiver. Receiving 6LoRa devices consume
approximately 4 times less energy (2.93\,mW) than class C devices (12.41\,mW) due to
the transceiver being active approximately 25\% of the time (one slot for beacon reception and
three slots for GTS reception).

In conclusion, our findings validate that class A devices demonstrate the lowest
power consumption, while class C devices exhibit the highest power
consumption. Although 6LoRa is viable for battery-powered
devices, the power consumption is more than 3 times higher than that of class A devices.

\section{Conclusions and outlook}\label{sec:conclusions}

In this work,  we proposed a system architecture (6LoRa) that enables 
IPv6 communication for the DSME MAC over LoRa. This overcomes the limitations of LoRaWAN,
a widely used LPWAN technology.
The evaluation on common off-the-shelf devices confirmed that our system
renders better communication performance for peer-to-peer scenarios than
SCHC-LoRaWAN. Our solution requires more memory resources than SCHC-LoRaWAN,
although it remains compatible with constrained-devices. The proposed approach
facilitates energy-efficient IPv6 communication over LoRa, obviating the need
for additional backhaul infrastructure, which stands in contrast to the LoRaWAN
network paradigm.

There are three future directions for this work. First, studying allocation strategies
DSME-LoRa networks may potentially facilitate real-time communication over LoRa.
Second, optimizing its operation on battery-powered devices,
aiming to achieve deterministic wireless communications at high energy
efficiency. Third, the evaluation of SCHC compression over
DSME-LoRa shall open new research possibilities.

%
%

%

%
%

%
%
\balance
\bibliographystyle{abbrv}
\bibliography{own,rfcs,ids,ngi,iot,layer2,meta,complexity,internet,theory}
\end{document}